\documentclass[final,5p]{elsarticle}

\usepackage[T1]{fontenc}
\usepackage[utf8]{inputenc}
\usepackage{microtype}
\usepackage{amsmath,amssymb}
\usepackage{mathtools}
\usepackage{booktabs}
\usepackage{multirow}
\usepackage{graphicx}
\usepackage{subcaption}
\usepackage{xcolor}
\usepackage{textcomp}
\usepackage{algorithm}
\usepackage{algorithmicx}
\usepackage{algpseudocode}
\usepackage{listings}
\definecolor{codekeyword}{rgb}{0.0, 0.4, 0.65}
\definecolor{codecomment}{rgb}{0.4, 0.45, 0.5}
\definecolor{codestring}{rgb}{0.2, 0.5, 0.25}
\definecolor{codenumber}{rgb}{0.75, 0.75, 0.75}
\definecolor{codebase}{rgb}{0.15, 0.15, 0.15}
\usepackage{pgfplots}
\usepgfplotslibrary{fillbetween}
\usepgfplotslibrary{groupplots}
\pgfplotsset{compat=1.18}
\pgfplotsset{legend image code/.code={\draw[#1] (0cm,0cm) -- (0.4cm,0cm);}}
\usetikzlibrary{spy}
\usepackage{siunitx}
\DeclareSIUnit{\lups}{LUP/s}
\DeclareSIUnit{\flop}{FLOP}
\definecolor{lyrA}{rgb}{0.20, 0.50, 0.25}
\definecolor{lyrB}{rgb}{0.02, 0.47, 0.45}
\definecolor{lyrC}{rgb}{0.00, 0.40, 0.65}
\definecolor{lyrD}{rgb}{0.30, 0.28, 0.58}
\usepackage{tikz}
\pgfplotsset{compat=newest}
\definecolor{pv1}{rgb}{0.278431, 0.278431, 0.858824}
\definecolor{pv2}{rgb}{0, 0, 0.360784}
\definecolor{pv3}{rgb}{0, 1, 1}
\definecolor{pv4}{rgb}{0, 0.501961, 0}
\definecolor{pv5}{rgb}{1, 1, 0}
\definecolor{pv6}{rgb}{1, 0.380392, 0}
\definecolor{pv7}{rgb}{0.419608, 0, 0}
\definecolor{pv8}{rgb}{0.878431, 0.301961, 0.301961}
\pgfplotsset{
    colormap={papercmap}{
        rgb255(0cm)=(29, 100, 171)
        rgb255(1cm)=(255, 255, 255)
        rgb255(2cm)=(218, 66, 40)
    }
}

\usepackage{hyperref}
\hypersetup{colorlinks=true, citecolor={blue!55!black}, linkcolor={blue!55!black}, urlcolor={blue!55!black}}

\usepackage{atbegshi}
\newcounter{fnqshipout}
\AtBeginShipout{\stepcounter{fnqshipout}%
  \ifnum\value{fnqshipout}=1 \hoffset=0.8cm \else \hoffset=0pt \fi}
\begin{document}

\begin{frontmatter}

\title{A Gradient-Reconstruction Lattice Boltzmann Method for Compressible Navier--Stokes--Fourier Equations}

\author[lbrg,ianm]{Adrian Kummerländer\corref{cor1}\fnref{eq}}
\ead{kummerlaender@kit.edu}
\author[lbrg,mvm]{Fedor Bukreev\fnref{eq}}
\author[lbrg,ianm,mvm]{Mathias J. Krause\fnref{eq}}
\cortext[cor1]{Corresponding author.}
\fntext[eq]{These authors contributed equally to this work.}
\affiliation[lbrg]{organization={Lattice Boltzmann Research Group (LBRG), Karlsruhe Institute of Technology (KIT)}, city={Karlsruhe}, postcode={76131}, country={Germany}}
\affiliation[ianm]{organization={Institute for Applied and Numerical Mathematics (IANM), Karlsruhe Institute of Technology (KIT)}, city={Karlsruhe}, postcode={76131}, country={Germany}}
\affiliation[mvm]{organization={Institute of Mechanical Process Engineering and Mechanics (MVM), Karlsruhe Institute of Technology (KIT)}, city={Karlsruhe}, postcode={76131}, country={Germany}}

\begin{abstract}
Reaching compressible flow has usually forced lattice Boltzmann methods to abandon the compact stencil or the strict locality that make them efficient.
We give up neither, solving the compressible Navier--Stokes--Fourier equations with a scheme that transports only the conserved mass, momentum and energy.
The viscous stress and heat flux depend on gradients of the velocity and temperature.
To capture them, existing compressible schemes go beyond a single lattice of the conserved fields.
They enlarge the velocity set, carry the stress and heat flux as extra transported fields, place the energy on a separate grid, or give up exact streaming for an off-lattice advection.
We instead recover the gradients from the non-equilibrium part of the distributions already present, inside the collision.
No field beyond the conserved state is transported, no neighbour is read, streaming stays exact, and every computation runs in single precision.
The method carries five fields where a transported-flux scheme carries fourteen, at a fraction of the memory and $4.9$ times the throughput.
Verified against the exact Sod and Becker shock solutions and a supersonic Taylor--Green vortex, it tracks the direct-numerical-simulation reference on the kinetic energy and stays within the reference-solver spread on both dissipation rates.
\end{abstract}

\begin{keyword}
lattice Boltzmann method \sep compressible Navier--Stokes--Fourier \sep shock capturing \sep high-performance computing
\end{keyword}

\end{frontmatter}

\section{Introduction}
The lattice Boltzmann method owes its efficiency to two structural properties: exact streaming along a compact, fixed set of lattice links, and a collision that touches only the local cell~\cite{krueger2017a}.
Reaching compressible flow has meant giving one of these up.
High-order Hermite equilibria and enlarged or multi-speed velocity sets~\cite{shan2006, coreixas2017, frapolli2016, atif2025} widen the stencil.
Semi-Lagrangian off-lattice schemes~\cite{dorschner2018, wilde2020, wilde2021} abandon exact on-lattice streaming.
Hybrid formulations~\cite{feng2019, jacob2018, guo2020, renard2021, vienne2025} close the energy equation with a finite-difference or finite-volume solver alongside the lattice.

The extended-state route keeps both properties at the cost of transporting more state.
It assigns a lattice to each component of an extended macroscopic state~\cite{bukreev2026b, guillon2024, wissocq2025, hosseini2026a}, a discrete-kinetic relaxation approximation of conservation laws~\cite{jin1995, natalini1998, aregbadriollet2000, bouchut1999}.
The automatic construction of~\cite{kummerlaender2026d} turns such a declared system into a lattice Boltzmann scheme targeting OpenLB~\cite{krause2021, kummerlaender2023, kummerlaender2026c}.
For the \emph{Compressible Navier--Stokes--Fourier} (CNSF) equations the relaxed-flux scheme~\cite{bukreev2026a} carries the viscous stress as a symmetric tensor state and the heat flux as a vector state.
These are nine fields beyond the five conserved ones: fourteen fields and ninety-eight streamed populations per cell on the D3Q7 lattice.
Since lattice Boltzmann is commonly bound by memory bandwidth rather than arithmetic~\cite{krueger2017a, tolke2010, januszewski2014, godenschwager2013, bauer2021b, bauer2021a}, this transported state sets the achievable throughput.
The stress and heat flux are auxiliary: they supply the velocity and temperature gradients the Newtonian and Fourier closures require, yet the conserved state determines them.
Supplying them is nonetheless the hard part, one a recent vectorial compressible solver judges difficult enough to stay inviscid~\cite{hosseini2026}, because its leading-order numerical dissipation is already of the order of the physical stress under acoustic scaling.
The full dissipative system has already been solved by transporting that state, from magnetohydrodynamics~\cite{bukreev2026b} to the Navier--Stokes--Fourier equations~\cite{bukreev2026a}, and the reconstruction operator~\cite{kummerlaender2026e} recovers the same physical stress and heat flux from the non-equilibrium moment regardless.

The gradients the closures require are already present in the conserved fields' non-equilibrium moments, and recovering them during collision yields a compressible scheme that transports the five conserved fields alone.
To our knowledge it is the first to reach supersonic, shocklet-laden Navier--Stokes--Fourier turbulence with exact streaming on a nearest-neighbour (D3Q7) lattice and no auxiliary state.

The reconstruction rests on a general result, developed and verified by manufactured solutions in the supporting methodology paper~\cite{kummerlaender2026e}: to leading order the first moment of a conserved field's non-equilibrium part carries the gradient of that field, and an operator built from the diffusive-flux Jacobian recovers it exactly.
This paper realizes that result for the compressible Navier--Stokes--Fourier system and studies the scheme it yields: the coupled Newtonian-stress and Fourier-conduction Jacobian, the Courant-limited timestep, the validation against a \emph{Direct Numerical Simulation} (DNS) reference, and the memory and throughput of transporting only the conserved fields.
The reconstruction collapses the fourteen fields to five and the ninety-eight streamed populations to thirty-five, forming the Newtonian stress and Fourier flux during the collision instead of transporting them as relaxed states.

The contributions are:
\begin{itemize}
\item a lattice Boltzmann method for the compressible Navier--Stokes--Fourier equations that reconstructs the velocity and temperature gradients inside the collision, reaching supersonic shocklet-laden turbulence on the compact nearest-neighbour D3Q7 lattice with exact streaming
\item its closed-form operator, built from the diffusive-flux Jacobian that couples the Newtonian stress and Fourier conduction
\item an explicit scheme with a single operating point, stable up to a Courant limit near $0.7$ and transferring unchanged across the validation cases
\item validation against exact shock solutions and a supersonic DNS turbulence reference, staying within its band throughout
\item the most accurate second-order scheme in that comparison, ahead of the only other one in every channel
\item numerical robustness sufficient to run entirely in single precision
\item a minimal transported state that runs at $4.9$ times the throughput of the fourteen-field relaxed-flux scheme and reaches $512^3$ on a two-GPU node
\end{itemize}

We evaluate the scheme on the compressible benchmarks of~\cite{bukreev2026a}.
The Sod problem~\cite{sod1978} tests captured discontinuities against an exact Riemann solution.
Becker's traveling wave~\cite{becker1922, morduchow1949} tests the coupled stress and heat-flux closure against a smooth closed-form profile.
The compressible Taylor--Green vortex~\cite{taylor1937, brachet1983} at $\mathrm{Re}=1600$ and $M_0=1.25$ combines both, measured against the seven-solver comparison and $2048^3$ reference of Chapelier et al.~\cite{chapelier2024}.
Section~\ref{sec:methodology} states the system, its discretization on the D3Q7 lattice and the gradient reconstruction, Section~\ref{sec:validation} presents the validation, and Section~\ref{sec:performance} the computational efficiency.

\section{Methodology}\label{sec:methodology}

\subsection{Governing Equations}
The method of~\cite{kummerlaender2026d} takes a system of conservation laws as its entry point,
\begin{equation}
\partial_t \mathbf{Q} + \nabla\cdot\boldsymbol{\Phi} = \mathbf{S},
\end{equation}
with $\mathbf{Q}$ the conserved state, $\boldsymbol{\Phi}$ the physical flux and $\mathbf{S}$ a local source.
The compressible viscous gas is the system of the relaxed-flux scheme~\cite{bukreev2026a}, taken here in its conserved form,
\begin{equation}\label{eq:system}
\mathbf{Q} = \begin{pmatrix*}[l] \rho \\ \rho\mathbf{u} \\ E \end{pmatrix*}, \quad
\boldsymbol{\Phi} = \begin{pmatrix*}[l] \rho\mathbf{u} \\ \rho\mathbf{u}\otimes\mathbf{u} + p\mathbf{I} - \boldsymbol{\tau} \\ (E+p)\mathbf{u} - \boldsymbol{\tau}\cdot\mathbf{u} + \mathbf{q} \end{pmatrix*}, \quad
\mathbf{S} = \mathbf{0},
\end{equation}
with the Newtonian stress $\boldsymbol{\tau} = \mu(\nabla\mathbf{u} + \nabla\mathbf{u}^{\mathsf{T}} - \tfrac{2}{3}(\nabla\cdot\mathbf{u})\mathbf{I})$ and the Fourier flux $\mathbf{q} = -\kappa\nabla T$.
The pressure is $p = (\gamma-1)(E - \tfrac{1}{2}\rho|\mathbf{u}|^2)$, the temperature $T = p/\rho$, and the dynamic viscosity $\mu(T)$ follows the Sutherland law of the benchmark~\cite{chapelier2024}.
The relaxed-flux scheme carries $\boldsymbol{\tau}$ and $\mathbf{q}$ as nine additional states relaxed toward these closures~\cite{bukreev2026a, jin1995}.
Here they are the closures themselves, evaluated from the gradients reconstructed in Section~\ref{sec:reconstruction}, so the transported state is the five conserved fields alone.

\subsection{Discretization}
From this declaration a lattice Boltzmann scheme is constructed by the procedure of~\cite{kummerlaender2026d, kummerlaender2026e}.
Each conserved field carries an independent set of populations $f_{k,i}$.
Its flux $\boldsymbol{\Phi}_k$ sits in the first moment of a linear equilibrium, and $Q_k$ is recovered from the zeroth moment.
The scheme marches by a collide-and-stream update at per-field relaxation frequency $\omega_k = 1/\tau_{\mathrm{LB},k}$, with the source projected onto the lattice weights.
This work uses two collision operators.
The manufactured solution and the subsonic vortex are smooth and use BGK.
The two shock tubes and the supersonic Taylor--Green vortex use the \emph{Regularized Lattice Boltzmann} (RLB) operator of Section~\ref{sec:shock_capturing}, which the compression sensor activates in place of BGK.
Multiple-relaxation operators are available from the same declaration but unused.
The D3Q7 stencil used here has $c_s^2 = 1/4$, so the five fields carry thirty-five populations per cell.
Every result reported here uses single-precision populations under an equilibrium shift, storing the deviation from a uniform reference state~\cite{kummerlaender2026d}, which keeps the significant bits on the fluctuations rather than the mean.

\subsection{Gradient Reconstruction}\label{sec:reconstruction}
The Newtonian stress and Fourier flux of~\eqref{eq:system} depend on first gradients of the state: the stress on $\nabla\mathbf{u}$, the heat flux on $\nabla T$.
The relaxed-flux scheme~\cite{bukreev2026a} supplies them by transporting the stress and heat flux as nine relaxed states.
Here they are instead recovered inside the collision from the conserved fields' non-equilibrium first moments, by the gradient reconstruction of~\cite{kummerlaender2026e}, so no auxiliary state is transported.
The full declaration is given in Listing~\ref{lst:declaration}~\cite{kummerlaender2026d}.

By the Chapman--Enskog expansion the non-equilibrium first moment of field $k$ carries its gradient to leading order~\cite{kummerlaender2026e}.
Subtracting the advective flux $\boldsymbol{\Phi}_k^{\mathrm{adv}} = \boldsymbol{\Phi}_k|_{\nabla\mathbf{Q}=\mathbf{0}}$, a function of the conserved moments alone, gives the direct reconstruction
\begin{equation}\label{eq:gradrecon}
  \nabla\widetilde{Q}_k = -\frac{1}{c_s^2\,\tau_{\mathrm{LB},k}}\,\boldsymbol{\Pi}_k^{(1)}, \qquad
  \boldsymbol{\Pi}_k^{(1)} = \sum_i f_{k,i}\,\boldsymbol{\xi}_i - \boldsymbol{\Phi}_k^{\mathrm{adv}},
\end{equation}
with both terms known during the collision.
The moment still carries the diffusive flux $\boldsymbol{\Phi}^{\mathrm{diff}} = \mathbf{M}\,\nabla\mathbf{Q}$, with $\mathbf{M} = \partial\boldsymbol{\Phi}^{\mathrm{diff}}/\partial(\nabla\mathbf{Q})$ the diffusive-flux Jacobian, so the true gradient follows by the reconstruction operator
\begin{equation}\label{eq:feedback}
  \nabla\mathbf{Q} = \mathbf{R}\,\nabla\widetilde{\mathbf{Q}}, \qquad
  \mathbf{R} = \left(\mathbf{I} - c_s^{-2}\,\mathrm{diag}(\omega_k)\,\mathbf{M}\right)^{-1},
\end{equation}
assembled in closed form from the declared closures~\cite{kummerlaender2026e}.
The primitive gradients $\nabla\mathbf{u}$ and $\nabla T$ that the closures need then follow from the reconstructed conserved gradients by the chain rule, evaluated in registers alongside $\mathbf{R}$, whose compressible block structure is set out in Section~\ref{sec:cnsf_operator}.
The reconstruction is explicit: the gradients respond within the step where the relaxed states of~\cite{bukreev2026a} lag by a relaxation time.
The stable timestep is bounded by a Courant ceiling of order $\mathrm{CFL} \approx 0.7$ (Section~\ref{sec:numerical_parameters}).

\begin{lstlisting}[language=Python, float=htbp, label={lst:declaration},
caption={The compressible Navier--Stokes--Fourier system as declared to the compiler. Only the five
conserved fields are declared, and \lstinline|reconstruct_gradients| recovers $\nabla\mathbf{u}$ and
$\nabla T$ in-collide from their non-equilibrium first moments. The fourteen-field form
of~\cite{bukreev2026a} instead declares the viscous stress and heat flux as additional transported
states.}]
from pde2lbm import *
eqs = ConservationLaws(dim=3, lattice="D3Q7")

# conserved fields
rho  = eqs.state("rho",  dim=mass / length**3)
rhou = eqs.state("rhou", shape=3,
                 dim=momentum / length**3)
E    = eqs.state("E",    dim=energy / length**3)

# parameters
gamma = eqs.parameter("gamma")
mu    = eqs.parameter("mu",    dim=pressure * time)
kappa = eqs.parameter("kappa", dim=pressure * time)
T_ref = eqs.parameter("T_ref", dim=energy / mass)

# primitives
u    = rhou / rho
p    = (gamma - 1) * (E - 0.5 * rho * u.dot(u))
T    = p / rho

# Sutherland temperature-dependent viscosity
T_S  = 0.4042
mu_T = (mu * (1 + T_S)
        * (T / T_ref)**Rational(3, 2)
        / (T / T_ref + T_S))

# gradient-form closures: stress and flux
tau  = mu_T * (grad(u) + grad(u).T
               - 2/3 * div(u) * eye(3))
q    = -kappa * (mu_T / mu) * grad(T)

# conservation laws
eqs.add([
    Eq(dt(rho)  + div(rhou), 0),
    Eq(dt(rhou)
       + div(outer(rhou, u) + p * eye(3) - tau),
       zeros(3, 1)),
    Eq(dt(E)    + div((E + p) * u - tau * u + q), 0),
])

# recover gradients from the non-eq moment
eqs.reconstruct_gradients([u, T])
# default collision is BGK; the shock variant
# selects RLB and compiles in Listing 2
\end{lstlisting}

\subsection{The Compressible Reconstruction Operator}\label{sec:cnsf_operator}

The construction of~\cite{kummerlaender2026e} builds the operator $\mathbf{R}$ of~\eqref{eq:feedback} automatically from the declared flux, through the diffusive-flux Jacobian $\mathbf{M} = \partial\boldsymbol{\Phi}^{\mathrm{diff}}/\partial(\nabla\mathbf{Q})$ over the conserved gradients $(\nabla\rho, \nabla(\rho\mathbf{u}), \nabla E)$.
This section exhibits the generated $\mathbf{M}$ and $\mathbf{R}$ for the CNSF system row by row.
The continuity flux is purely advective, so the mass row of $\mathbf{M}$ vanishes and $\nabla\rho$ passes through the operator unchanged.
The momentum row is the Newtonian block of~\cite{kummerlaender2026e}, with $\mathsf{S}$ the symmetric-deviatoric map.
By the chain rule $\nabla\mathbf{u} = (\nabla(\rho\mathbf{u}) - \mathbf{u}\otimes\nabla\rho)/\rho$ the stress $-\boldsymbol{\tau}$ acts on $\nabla(\rho\mathbf{u})$ through $-(\mu/\rho)\,\mathsf{S}$ and on $\nabla\rho$ through the same map weighted by the velocity.
The energy row carries the compressible content.
Its diffusive flux $-\boldsymbol{\tau}\cdot\mathbf{u} + \mathbf{q}$ contracts the stress block with $\mathbf{u}$ as viscous work and adds Fourier conduction, whose temperature gradient expands through $T = (\gamma-1)(E - \tfrac{1}{2}\rho|\mathbf{u}|^{2})/\rho$ as
$\nabla T = \tfrac{\gamma-1}{\rho}\big(\nabla E - \sum_j u_j\,\nabla(\rho u_j)\big) + \tfrac{1}{\rho}\big(\tfrac{\gamma-1}{2}|\mathbf{u}|^{2} - T\big)\nabla\rho$,
so conduction reads all three conserved gradients.
Collecting rows, with $(\nabla\mathbf{a})_{jc} = \partial_c a_j$ so that the transpose contracts the field index,
\begin{figure*}[tbp]
\begin{equation}\label{eq:cnsfM}
  \mathbf{M} =
  \begin{pmatrix}
    \mathbf{0}
    & \mathbf{0}
    & \mathbf{0} \\[8pt]
    \dfrac{\mu}{\rho}\,\mathsf{S}\big(\mathbf{u}\otimes\,\cdot\,\big)
    & -\dfrac{\mu}{\rho}\,\mathsf{S}
    & \mathbf{0} \\[12pt]
    \dfrac{\mu}{\rho}\Big(|\mathbf{u}|^{2}\,\mathbf{I} + \tfrac{1}{3}\,\mathbf{u}\otimes\mathbf{u}\Big)
    + \dfrac{\kappa}{\rho}\Big(T - \tfrac{\gamma-1}{2}|\mathbf{u}|^{2}\Big)\mathbf{I}
    & -\dfrac{\mu}{\rho}\,\big(\mathsf{S}\,\cdot\,\big)\,\mathbf{u}
    + \dfrac{(\gamma-1)\,\kappa}{\rho}\,(\,\cdot\,)^{\mathsf{T}}\mathbf{u}
    & -\dfrac{(\gamma-1)\,\kappa}{\rho}\,\mathbf{I}
  \end{pmatrix}
\end{equation}
\end{figure*}
acting on the column $(\nabla\rho, \nabla(\rho\mathbf{u}), \nabla E)^{\mathsf{T}}$.
In the field ordering $(\rho, \rho\mathbf{u}, E)$ the Jacobian is therefore block lower triangular: mass feeds momentum, both feed energy, and no flux reads $\nabla E$ except conduction itself.
The inverse $\mathbf{R}$ inherits this pattern and is evaluated by a forward block solve.
The momentum block is that of the incompressible case~\cite{kummerlaender2026e}: $\mathsf{S}$ annihilates the antisymmetric and trace parts, so vorticity and dilatation pass through undamped and only the symmetric-deviatoric strain is scaled by $1/(1+\beta)$.
The lattice viscous number
\begin{equation}\label{eq:beta}
  \beta = \frac{2\,\nu_{\mathrm{lat}}\,\omega}{c_s^{2}},
\end{equation}
with $\nu_{\mathrm{lat}}$ the viscosity in lattice units, sets how far the operator departs from the identity.
Within this block the shear pairs and the normal-stress components are mutually coupled and are inverted together.
The energy row then reduces to a scalar inversion with the corrected upstream gradients inserted on its right side,
\begin{equation}\label{eq:cnsfE}
\begin{gathered}
  \nabla E = \frac{\nabla\widetilde{E} + c_s^{-2}\,\omega_E\left[\mathbf{M}_{E\rho}\,\nabla\rho + \mathbf{M}_{Em}\,\nabla(\rho\mathbf{u})\right]}{1 + c_s^{-2}\,\omega_E\,\gamma\,\alpha_{\mathrm{lat}}}, \\[4pt]
  \alpha = \frac{\kappa}{\rho\,c_p}, \qquad c_p = \frac{\gamma}{\gamma-1},
\end{gathered}
\end{equation}
with $\mathbf{M}_{E\rho}$ and $\mathbf{M}_{Em}$ the energy-row blocks of~\eqref{eq:cnsfM}, $\alpha$ the thermal diffusivity and $\alpha_{\mathrm{lat}}$ its value in lattice units.
The factor $\gamma\alpha$ appears because the conducted quantity is the total energy while $\kappa$ conducts temperature.
The conduction damping is thus the same $1/(1+\cdot)$ reduction as the shear block with $2\nu$ replaced by $\gamma\alpha$: for equal relaxation frequencies the two coefficients stand in the ratio $\gamma/(2\,\mathrm{Pr})$ with the Prandtl number $\mathrm{Pr} = \mu\,c_p/\kappa$.
At leading order in the lattice Mach number the surviving entries are the two diagonal damping factors and the conduction weight on $\nabla\rho$, whose coefficient $\kappa T/\rho$ does not vanish with the velocity.
The viscous-work couplings and the Fourier weight on $\nabla(\rho\mathbf{u})$ carry a factor $\mathbf{u}$ and enter at $\mathcal{O}(\mathrm{Ma})$, the stress weight on $\nabla\rho$ in the energy row at $\mathcal{O}(\mathrm{Ma}^{2})$.
Every entry of $\mathbf{M}$ depends on the state, through $\rho$, $\mathbf{u}$ and the Sutherland viscosity $\mu(T)$, so $\mathbf{R}$ is assembled per cell inside the collision~\cite{kummerlaender2026e}.
The magnitude of the correction depends on the working point.
At the compressible Taylor--Green regime ($\mathrm{Re} = 1600$, Section~\ref{sec:ctgv}) the physical viscosity is small, so $\beta \approx 0.02$ and $\mathbf{R}$ is within a few percent of the identity.
At the more viscous manufactured-solution and shock-tube working points $\beta$ grows with resolution from $0.5$ at $N = 64$ to $4$ at $N = 512$, so the operator is a substantial correction rather than a near-identity.

\subsection{Shock Capturing}\label{sec:shock_capturing}
The reconstruction supplies the physical dissipation of the resolved flow, but under-resolved shocks need extra damping.
Shock capturing is added with a local compression sensor, in the spirit of the relaxed-flux scheme~\cite{bukreev2026a}.
It forms a per-cell indicator $\chi$ from the compression $-\nabla\cdot\mathbf{u}$ already reconstructed within the step.
The indicator is nonzero only in compression and above a threshold $J_{\min}$, so smooth regions are left at the physical rate.
Established sensors instead read a neighborhood: the Jameson switch differences the pressure twice~\cite{jameson1981}, the Ducros gate compares dilatation against vorticity magnitude~\cite{ducros1999}, and localized artificial diffusivity smooths a gradient indicator over a stencil~\cite{cook2005, kawai2008}.
Here the dilatation is read from the reconstructed gradient, so the sensor is cell-local and needs no separate pass over the grid.

The collision operator is a choice of the declaration~\cite{kummerlaender2026d}, and the shock cases exercise it.
The sensor is a kernel extension that runs after the gradient reconstruction and before the collision, and the conserved fields then collide with a regularized operator in place of the default BGK (Listing~\ref{lst:shocksensor}).
The regularized collision reconstructs the non-equilibrium from its first moment and discards the higher moments the D3Q7 lattice does not constrain, and where $\chi$ is nonzero it lowers the relaxation rate of the reconstructed part to $\omega_h = \omega_k - \chi(\omega_k - \omega_{\min})$.
The two effects counter the near-$\omega=2$ operating point: the regularization removes an odd-even mode the discarded moments would otherwise carry through the smooth flow, and the reduced rate supplies the front dissipation the over-relaxed collision lacks.
The collision and the gradient reconstruction are emitted from the declaration and reproduced byte-for-byte when it is regenerated, so the executed kernel is exactly what the declaration specifies.

\begin{lstlisting}[language=Python, float=htbp, label={lst:shocksensor},
caption={Shock capturing as a choice of collision operator on the declaration of
Listing~\ref{lst:declaration}. The sensor writes the clamped compression indicator $\chi$ from the
reconstructed dilatation, and \lstinline|RegularizedShockLB| lowers the relaxation rate toward
$\omega_{\min}$ where $\chi$ fires, leaving smooth regions ($\chi=0$) at the unchanged rate.}]
# continues Listing 1 (same eqs)
gain      = eqs.parameter("SENSOR_GAIN")
J_min     = eqs.parameter("J_MIN")
omega_min = eqs.parameter("OMEGA_FLOOR")
chi       = eqs.input_field("chi")

class ShockSensor(KernelExtension):
    stage = "post_reconstruction"
    def emit(self, c):
        div_u = div(u)   # dilatation
        s     = -div_u / c_s   # compression Mach
        return store(chi, where(
            (div_u < 0) & (s > J_min),
            clamp(0, 1, gain * s), 0))

class RegularizedShockLB(RLB):
    # RLB with omega reduced toward omega_min
    def relaxation_rate(self, omega):
        return omega - chi * (omega - omega_min)

eqs.extend(ShockSensor())
for q in [rho, rhou[0], rhou[1], rhou[2], E]:
    eqs.set_collision(q, RegularizedShockLB())

eqs.compile(class_name="CompressibleNSF")
\end{lstlisting}

Algorithm~\ref{alg:collide} collects the complete per-cell update.
The reconstruction, the sensor and the positivity floor each read only the cell they act on, so the whole step is cell-local and the stress and heat flux never leave registers.

\begin{algorithm*}[!tb]
\caption{In-collide update of the five-field scheme, per cell and time step.}
\label{alg:collide}
\begin{algorithmic}[1]
\Require populations $f_{k,i}$, relaxation $\omega_k = 1/\tau_{\mathrm{LB},k}$, reconstruction operator $\mathbf{R}$ (Section~\ref{sec:cnsf_operator}), floors $\rho_{\min}, p_{\min}$, sensor gain $g$ and threshold $J_{\min}$, floor rate $\omega_{\min}$
\State $\mathbf{Q} \gets \big(\rho, \rho\mathbf{u}, E\big) = \textstyle\sum_i f_{k,i}$ \Comment{conserved moments}
\State $\rho \gets \max(\rho, \rho_{\min})$, \enspace $p \gets \max\!\big((\gamma-1)(E - \tfrac{1}{2}\rho|\mathbf{u}|^2), p_{\min}\big)$, \enspace $T \gets p/\rho$ \Comment{primitives, positivity floor}
\State $\boldsymbol{\Pi}^{(1)}_k \gets \textstyle\sum_i f_{k,i}\,\boldsymbol{\xi}_i - \boldsymbol{\Phi}^{\mathrm{adv}}_k(\mathbf{Q})$ \Comment{non-equilibrium first moment}
\State $\nabla\mathbf{Q} \gets \mathbf{R}\,\big(\!-\boldsymbol{\Pi}^{(1)}/(c_s^2\,\tau_{\mathrm{LB},k})\big)$, \enspace then $\nabla\mathbf{u}, \nabla T$ by the chain rule \Comment{reconstruct}
\State $s \gets -(\nabla\!\cdot\mathbf{u})/c_s$, \enspace $\chi \gets \mathrm{clamp}(0,1,\,g\,s)$ if $\nabla\!\cdot\mathbf{u}<0$ and $s>J_{\min}$, else $0$ \Comment{compression gate, dead-band $J_{\min}$}
\State $\boldsymbol{\tau} \gets \mu\big(\nabla\mathbf{u} + \nabla\mathbf{u}^{\mathsf{T}} - \tfrac{2}{3}(\nabla\!\cdot\mathbf{u})\mathbf{I}\big)$, \enspace $\mathbf{q} \gets -\kappa\,\nabla T$ \Comment{Newtonian and Fourier closures}
\State $f^{\mathrm{eq}}_{k,i} \gets f^{\mathrm{eq}}_{k,i}(\mathbf{Q}, \boldsymbol{\tau}, \mathbf{q})$ \Comment{closures enter the flux moment}
\State $\omega_h \gets \omega_k - \chi\,(\omega_k - \omega_{\min})$, \enspace reconstruct the first-moment part $r_{k,i}$ of $f_{k,i}-f^{\mathrm{eq}}_{k,i}$ \Comment{reduced rate, regularize}
\State $f_{k,i} \gets f^{\mathrm{eq}}_{k,i} + (1-\omega_h)\,r_{k,i} + w_i S_k$, \enspace then stream to $(\mathbf{x}+\mathbf{c}_i, t+1)$ \Comment{regularized collide and stream}
\end{algorithmic}
\end{algorithm*}

\subsection{Numerical Parameters}\label{sec:numerical_parameters}
The reconstruction leaves two numerical parameters free of the physics: the relaxation rate $\omega$ and the timestep, written as the Courant number $\mathrm{CFL} = \Delta t/\Delta x$.
Neither sets the physical transport: the viscosity and conductivity enter through the declared closure of Section~\ref{sec:cnsf_operator}, not through $\omega$, so both may be tuned for stability and accuracy without changing the modelled equations.

\begin{figure*}[!tb]
    \centering
    \includegraphics[width=\textwidth]{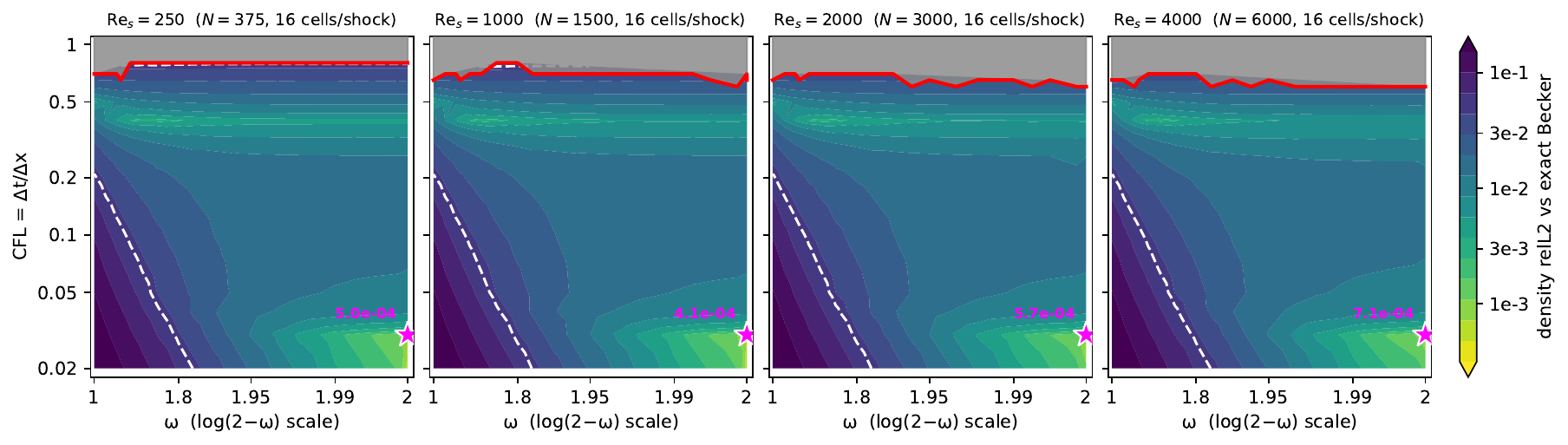}
    \caption{Operating map on the Becker viscous shock. Relative density $L_2$ error over the relaxation
    rate $\omega$ and the Courant number $\mathrm{CFL} = \Delta t/\Delta x$, at four shock Reynolds numbers.
    The $\omega$ axis uses a $\log(2-\omega)$ scale to resolve the over-relaxation edge. The red line is the
    stability ceiling, above which the run diverges (gray). The star marks each panel's minimum-error sample
    and the dashed line the $5\%$ accuracy limit.}
    \label{fig:becker_map}
\end{figure*}

Figure~\ref{fig:becker_map} maps their effect on the Becker viscous shock of Section~\ref{sec:shocktubes}, swept over $(\omega, \mathrm{CFL})$ at four shock Reynolds numbers $\mathrm{Re}_s$.
Each is resolved to the same shock thickness, $N = \tfrac{3}{2}\mathrm{Re}_s$ or about eight cells across the front, so the panels are directly comparable.
Two features fix the operating point.
Stability is bounded by a Courant ceiling of order $\mathrm{CFL} \approx 0.7$, drifting from about $0.8$ at $\mathrm{Re}_s = 250$ to about $0.65$ at $\mathrm{Re}_s = 4000$ over a sixteen-fold change in both $\mathrm{Re}_s$ and resolution.
Within the stable region the binding constraint is accuracy: the residual numerical viscosity vanishes as $\omega \to 2$, so the error falls toward the over-relaxation edge and is smallest there in every panel.

We therefore run the viscous cases with $\omega$ driven toward $2$ and $\mathrm{CFL}$ below the ceiling.
The shock-capturing parameters are fixed as in Section~\ref{sec:shock_capturing}.
Table~\ref{tab:run_params} collects the numerical parameters of every case.

\begin{table}[tb]
\centering
\caption{Numerical parameters of the validation cases. The manufactured solution and the Becker shock use
the dt-tied relaxation, so the $\omega$ shown is at the finest grid. Dashes mark the BGK cases, which run
without the sensor. $\gamma = 1.4$ and the positivity floors $\rho_{\min} = p_{\min} = 10^{-4}$ throughout.}
\label{tab:run_params}
\footnotesize
\setlength{\tabcolsep}{4pt}
\begin{tabular}{l c c c c c c}
\toprule
Case & Ma\,$^{a}$ & $\omega$ & CFL & $g$ & $J_{\min}$ & $\omega_{\min}$ \\
\midrule
MMS      & $2.0$  & $1.86$ & $0.05$\,$^{b}$            & --   & --          & --      \\
Sod      & $1.66$ & $1.40$ & $0.35$                    & $20$ & \num{3e-3}  & $1$     \\
Becker   & $2.0$  & $1.99$ & Fig.~\ref{fig:becker_map} & $20$ & \num{3e-3}  & $1$     \\
cTGV     & $1.25$ & $2.00$ & $0.1$                     & $40$ & \num{1e-2}  & $1.987$ \\
Subsonic & $0.5$  & $2.00$ & $0.1$                     & --   & --          & --      \\
\bottomrule
\end{tabular}
\\[3pt]
{\footnotesize $^{a}$ $M_0$ (MMS, cTGV, subsonic) or shock $M_s$ (Sod, Becker). \\
$^{b}$ acoustic timestep factor.}
\end{table}

\section{Validation}\label{sec:validation}
We verify the five-field scheme by manufactured solutions, the exact Sod~\cite{sod1978} and Becker~\cite{becker1922, morduchow1949} shock-tube solutions, and the supersonic compressible Taylor--Green vortex~\cite{chapelier2024}.
All cases run in single precision, which the manufactured-solution study confirms matches double precision to five significant figures on these grids.

\subsection{Manufactured Solution}\label{sec:mms}
Before the physical test cases we verify the order of accuracy of the gradient reconstruction on the compressible Navier--Stokes--Fourier system directly, by the method of manufactured solutions.
The manufactured source is formed automatically by the compiler's SymPy-based computer-algebra backend~\cite{kummerlaender2026d}.
It substitutes a chosen field into the continuous CNSF residual and differentiates symbolically, so no source is written by hand.
The collision reads $\nabla\mathbf{u}$ and $\nabla T$ from the non-equilibrium first moment $\boldsymbol{\Pi}^{(1)}$ within the step, by the reconstruction of Section~\ref{sec:reconstruction}.
The manufactured solution is a smooth compressible field at background Mach number $M_0 = 2$ and $\gamma = 1.4$ on the D3Q7 lattice ($c_s^2 = \tfrac14$).
On the unit periodic domain, with wavenumber $k = 2\pi$ and unit temporal period, the primitive fields are
\begin{equation}\label{eq:mms}
{\footnotesize
\begin{aligned}
\rho &= \rho_0\left[1 + 0.05\sin kx + 0.03\cos ky + 0.02\cos(kz+2\pi t)\right], \\
u_x &= c_s\left[a + 0.035\cos(kx+0.4) + 0.025\sin ky\right], \\
u_y &= c_s\left[a + 0.035\sin(ky+0.9) + 0.025\cos kz\right], \\
u_z &= c_s\left[a + 0.025\cos(kz+1.7) + 0.015\sin(kx+2\pi t)\right], \\
p &= p_0\big[1 + 0.05\cos(kx+1.2) + 0.03\sin(ky+2\pi t) \\
  &\qquad\quad {}+ 0.02\sin(kz+0.6)\big],
\end{aligned}
}
\end{equation}
with $a = 0.30/\sqrt{3}$ and the conserved energy $E = p/(\gamma-1) + \tfrac12\rho|\mathbf{u}|^2$.
The background $\rho_0$, $c_s$ and $p_0 = \rho_0\,c_s^2\,(0.30/M_0)^2/\gamma$ place the working point at background Mach $M_0 = 2$.
Density, pressure, and all three velocity components vary independently, so the velocity divergence, the temperature gradient, and the density gradient are nonzero everywhere.
This exercises the Newtonian stress with its bulk term, the Fourier flux, and every entry of the reconstruction operator.
The field is unsteady, so the full time-dependent system is verified rather than a steady snapshot.
Its amplitudes are kept small so that $\rho$ and $p$ stay positive and the velocity stays under the D3Q7 sub-characteristic ceiling, and its phases are distinct so that no gradient term cancels.
At this viscous working point the reconstruction operator is a substantial correction, with $\beta$ reaching $4$ (Section~\ref{sec:cnsf_operator}), rather than a near-identity, so the study tests the reconstruction itself.
The local Mach number ranges from $1.48$ to $2.54$, supersonic throughout, matching the regime of the physical test cases.
The field is refined acoustically to the final time $t = 1$ at fixed physical viscosity and conductivity, in the single-precision populations used throughout this paper.

At each resolution we report the relative $L_2$ error of each conserved field and take the \emph{Empirical Order of Convergence} (EOC) from the finest grid pair $N = 128 \to 256$, the coarser grids being pre-asymptotic.
All five conserved fields converge at second order (Table~\ref{tab:cnsf_mms}).
The single-precision result is indistinguishable from double precision to five significant figures at every resolution, so the discretization error, not the floating-point precision, sets the accuracy on these grids.

\begin{table}[tbp]
\centering
\footnotesize
\setlength{\tabcolsep}{4pt}
\caption{Manufactured-solution convergence of the five-field scheme on the three-dimensional compressible
Navier--Stokes--Fourier system at $M_0 = 2$. Relative $L_2$ error per conserved field and the empirical
order of convergence from the finest pair. Single and double precision agree, so one column is shown.}
\label{tab:cnsf_mms}
\begin{tabular}{l c c c c}
\toprule
Field & $N = 64$ & $N = 128$ & $N = 256$ & EOC \\
\midrule
$\rho$      & \num{4.133e-2} & \num{1.300e-2} & \num{3.456e-3} & $1.91$ \\
$\rho u_x$  & \num{6.418e-2} & \num{1.947e-2} & \num{5.128e-3} & $1.92$ \\
$\rho u_y$  & \num{7.038e-2} & \num{2.129e-2} & \num{5.603e-3} & $1.93$ \\
$\rho u_z$  & \num{4.829e-2} & \num{1.488e-2} & \num{3.940e-3} & $1.92$ \\
$E$         & \num{5.350e-2} & \num{1.652e-2} & \num{4.368e-3} & $1.92$ \\
\bottomrule
\end{tabular}
\end{table}

\subsection{Shock Tube}\label{sec:shocktubes}
\begin{figure*}[htbp]
    \centering
    \resizebox{\textwidth}{!}{\includegraphics{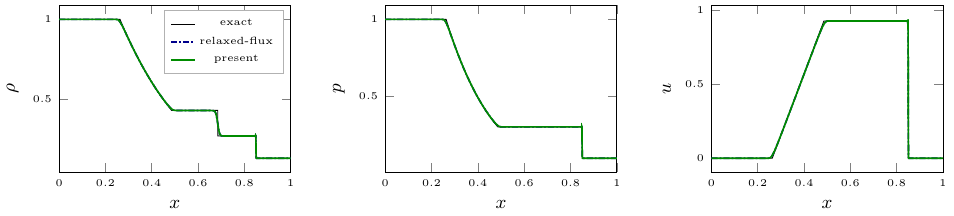}}
    \caption{Inviscid Sod shock tube at $N=3000$, over the full tube. Density, pressure and velocity
    against the exact Riemann solution~\cite{toro2009} (thin black), the fourteen-field relaxed-flux
    scheme of~\cite{bukreev2026a} (blue), and the present five-field reconstruction (green).}
    \label{fig:fneq_sod}
\end{figure*}

\begin{figure*}[htbp]
    \centering
    \resizebox{\textwidth}{!}{\includegraphics{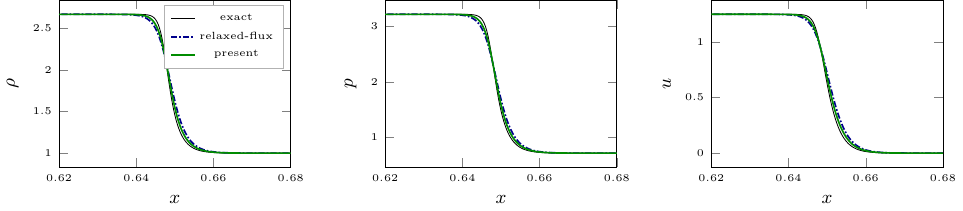}}
    \caption{Viscous Becker traveling wave at $N=3000$, zoomed to the shock region $x\in[0.62,0.68]$.
    Density, pressure and velocity against the exact traveling wave (thin black), the fourteen-field
    relaxed-flux scheme of~\cite{bukreev2026a} (blue), and the present five-field reconstruction (green).}
    \label{fig:fneq_becker}
\end{figure*}

Both shock tubes are evaluated against their exact solutions, the Riemann solution for Sod~\cite{toro2009} (Figure~\ref{fig:fneq_sod}) and the Becker traveling wave~\cite{becker1922, morduchow1949} for the viscous case (Figure~\ref{fig:fneq_becker}).

\paragraph{Configuration}
Both tubes use $\gamma = 1.4$ and the local non-equilibrium sensor of Section~\ref{sec:shock_capturing} at gain $g = 20$, $J_{\min} = 3\cdot10^{-3}$ and $\omega_{\min} = 1$.
The inviscid case runs at $\mu = 10^{-6}$ and $\mathrm{CFL} = 0.35$.
The viscous case uses the constant-$\mu$ Becker profile at $\mathrm{Pr} = 3/4$, with $\mu = 2\cdot10^{-3}$ from $M_s = 2$ and $\mathrm{Re}_s = 1000$ on unit scales.

\paragraph{Sod}
In the inviscid limit the reconstructed stress vanishes and the scheme reduces to the Euler flux, reproducing the fourteen-field relaxed-flux scheme.
The captured discontinuity converges sub-linearly in $L_2$ (Table~\ref{tab:sod_conv}).
The rate is bounded below unity, as it must be where a genuine discontinuity is spread over a fixed number of cells.
Density converges at order $\approx 0.3$, set by the contact discontinuity.
Pressure and velocity are smooth across the contact and reach order $0.5$ to $0.7$.
The profile is monotone with no overshoot.

\paragraph{Becker}
The viscous Becker shock exercises the reconstructed stress and heat flux, which vanish in the Sod limit.
The operating point follows Section~\ref{sec:numerical_parameters}: the relaxation rate is raised toward $\omega \to 2$ and the timestep kept below the Courant ceiling (Figure~\ref{fig:becker_map}).
Under acoustic refinement the dt-tied relaxation carries a residual numerical viscosity of order $\Delta x^2$, so $\omega$ approaches $2$ from below as the grid refines.
The shock converges toward second order.
The density error falls from $1.3\cdot10^{-1}$ at eight cells across the front to $1.6\cdot10^{-3}$ at $128$ cells (Table~\ref{tab:becker_conv}).
The observed order climbs to $1.9$ by $64$ cells, matching the manufactured-solution rate of Table~\ref{tab:cnsf_mms}.
That vanishing dissipation is what makes the sequence convergent.
At exactly $\omega = 2$ the collision adds none and the captured front is left undamped, so the finite-resolution errors lie above the $\omega = 2$ optima of the map and meet them as $\Delta x \to 0$.
The order softens at the finest step only as $\omega$ nears that edge.

\begin{table}[tbp]
    \centering
    \footnotesize
    \setlength{\tabcolsep}{4pt}
    \caption{Sod shock tube, relative $L_2$ error against the exact Riemann solution at $t=0.2$,
    trimmed to $x\in[0.05,0.95]$, with the base-2 order between successive $N$ in parentheses.}
    \label{tab:sod_conv}
    \resizebox{\columnwidth}{!}{\begin{tabular}{lccc}
\toprule
$N$ & $\rho\ L_2$ & $p\ L_2$ & $u\ L_2$ \\
\midrule
$750$  & \num{1.83e-2}         & \num{1.45e-2}         & \num{3.77e-2}         \\
$1500$ & \num{1.42e-2}\,(0.36) & \num{9.49e-3}\,(0.61) & \num{2.61e-2}\,(0.53) \\
$3000$ & \num{1.14e-2}\,(0.32) & \num{6.36e-3}\,(0.58) & \num{1.83e-2}\,(0.51) \\
$6000$ & \num{9.26e-3}\,(0.30) & \num{3.93e-3}\,(0.69) & \num{1.21e-2}\,(0.60) \\
\bottomrule
\end{tabular}
}
\end{table}

\begin{table}[tbp]
    \centering
    \footnotesize
    \setlength{\tabcolsep}{4pt}
    \caption{Viscous Becker shock ($\mathrm{Re}_s=1000$), relative $L_2$ error against the exact traveling
    wave, indexed by the cells resolving the front. The base-2 order is in parentheses.}
    \label{tab:becker_conv}
    \resizebox{\columnwidth}{!}{\begin{tabular}{lccc}
\toprule
cells/shock & $\rho\ L_2$ & $p\ L_2$ & $u\ L_2$ \\
\midrule
$8$   & \num{1.28e-1}         & \num{1.63e-1}         & \num{2.19e-1}         \\
$16$  & \num{4.53e-2}\,(1.50) & \num{5.79e-2}\,(1.49) & \num{7.12e-2}\,(1.62) \\
$32$  & \num{1.37e-2}\,(1.73) & \num{1.80e-2}\,(1.69) & \num{1.94e-2}\,(1.88) \\
$64$  & \num{3.69e-3}\,(1.89) & \num{5.00e-3}\,(1.85) & \num{5.15e-3}\,(1.91) \\
$128$ & \num{1.64e-3}\,(1.17) & \num{2.76e-3}\,(0.86) & \num{2.39e-3}\,(1.11) \\
\bottomrule
\end{tabular}
}
\end{table}

\subsection{Compressible Taylor--Green Vortex}\label{sec:ctgv}
\begin{figure*}[tbp]
    \centering
    \includegraphics[width=\textwidth]{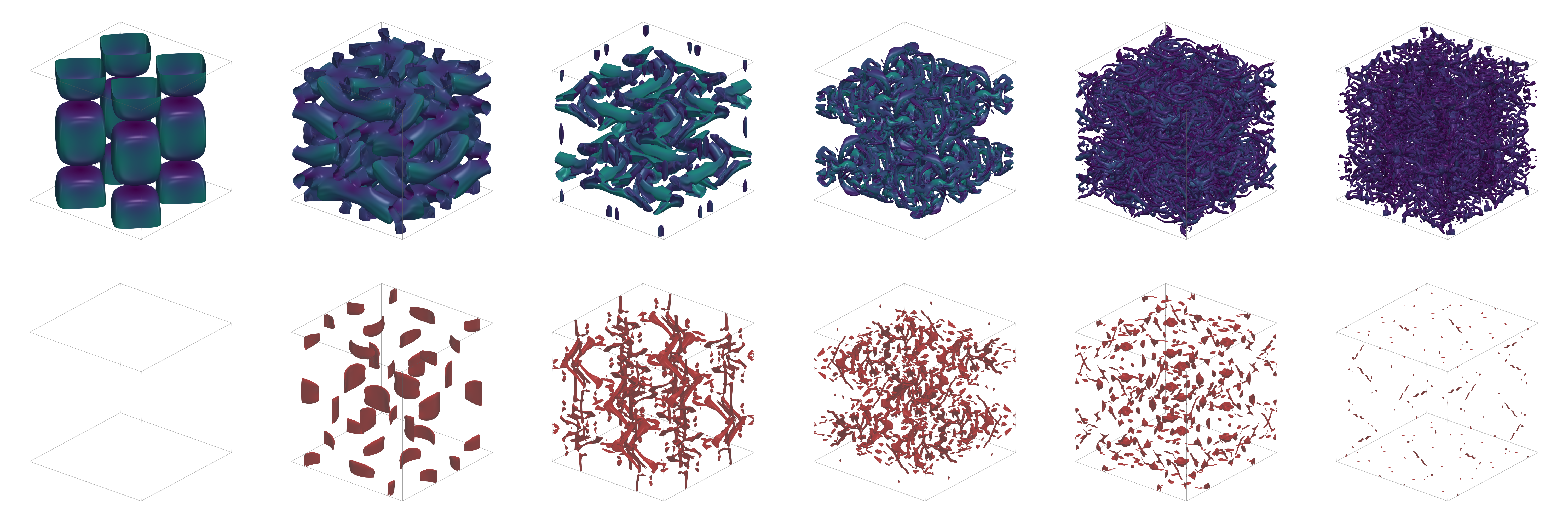}
    \caption{Flow visualization of the present $256^3$ result at $t_c \in \{0, 2.5, 5, 9, 14, 20\}$.
    Top, vortices as $Q$-criterion iso-surfaces colored by the local Mach number. Bottom, shocklets as
    iso-surfaces of the dilatation $\theta=\nabla\cdot\mathbf{u}$ with $\theta/\theta' < -3$~\cite{peng2018},
    $\theta'$ the r.m.s.\ dilatation. Shocklets form through transition and are sparse by $t_c=20$.}
    \label{fig:fneq_ctgv_viz}
\end{figure*}

\begin{figure*}[tbp]
    \centering
    \includegraphics{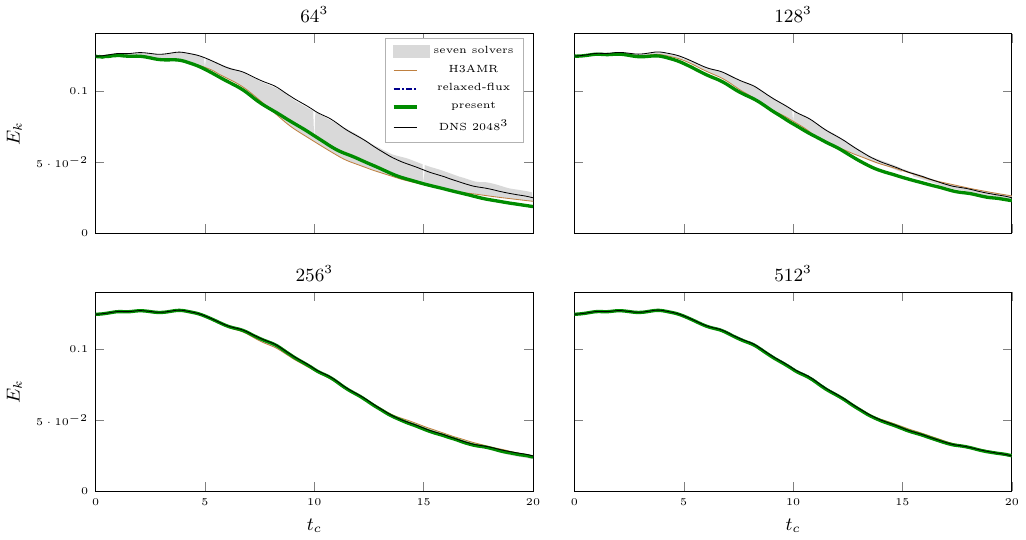}
    \caption{Compressible Taylor--Green vortex, $\mathrm{Re}=1600$, $M_0=1.25$. Time evolution of kinetic
    energy at four resolutions. The grey band is the spread of the seven solvers of
    Chapelier et al.~\cite{chapelier2024}, brown their H3AMR (the only second-order scheme). Blue is the
    fourteen-field relaxed-flux scheme of~\cite{bukreev2026a}, green the present five-field scheme, and thin black
    the $2048^3$ reference.}
    \label{fig:fneq_ctgv}
\end{figure*}

The compressible Taylor--Green vortex at $\mathrm{Re}=1600$, $M_0=1.25$ is the regime the reconstruction targets, under-resolved compressible turbulence with vortices and shocklets (the weak shocks embedded in the flow), shown in Figure~\ref{fig:fneq_ctgv_viz}.
Figure~\ref{fig:fneq_ctgv} shows the kinetic energy at four resolutions against the $2048^3$ \emph{Targeted Essentially Non-Oscillatory} (TENO)~\cite{fu2016} reference and the seven-solver comparison of Chapelier et al.~\cite{chapelier2024}, and Table~\ref{tab:ctgv_l2} gives the relative $L_2$ errors at $256^3$ and $512^3$.
Time is reported in convective units $t_c = t\,V_0/L_0$, with $V_0$ the initial peak velocity and $L_0$ the vortex length scale.
Errors are relative $L_2$ over $t_c\in[0,20]$ against the reference digitized from the published curves, computed identically for every curve.
Dissipation is computed spectrally throughout, matching the spectral post-processing of the reference solvers.
A finite-difference dilatation under-reads the shocklets and would bias the dilatational channel low.
The background dissipation rate $f_\mu$ sets a purely numerical damping on the collision, entering only the relaxation as $\tau = \tfrac12 + 4 f_\mu\,\nu\,\Delta t/\Delta x^2$ with $\nu = \mu/\rho_0$ the physical kinematic viscosity.
It is not a rescaling of the physical viscosity, which the reconstructed stress supplies in full, and is set per resolution, $f_\mu = 1.8$, $0.9$, $0.001$ and $0.01$ at $64^3$ through $512^3$.

\begin{figure*}[tbp]
    \centering
    \includegraphics{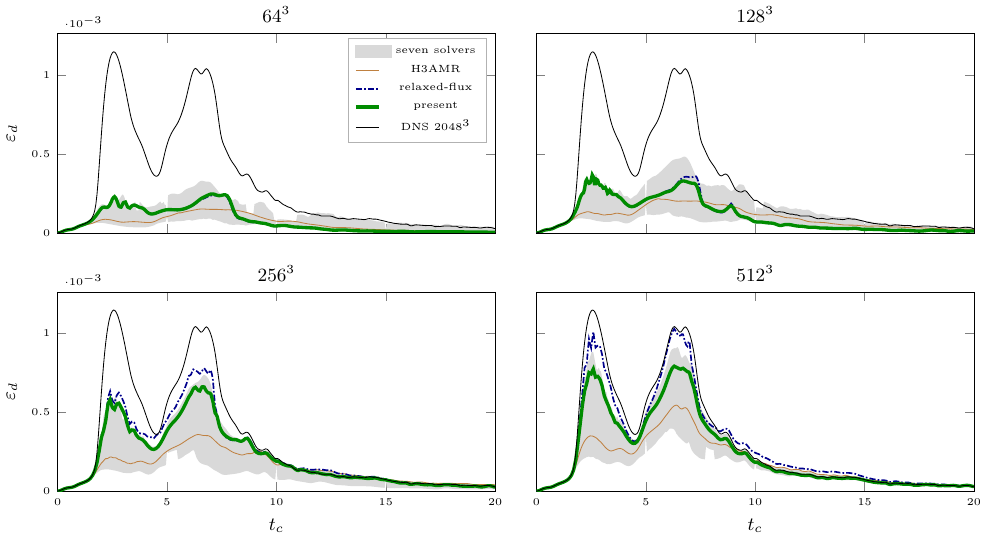}
    \caption{Time evolution of dilatational dissipation, curves as in Figure~\ref{fig:fneq_ctgv}.}
    \label{fig:fneq_ctgv_epsd}
\end{figure*}

\begin{figure*}[tbp]
    \centering
    \includegraphics{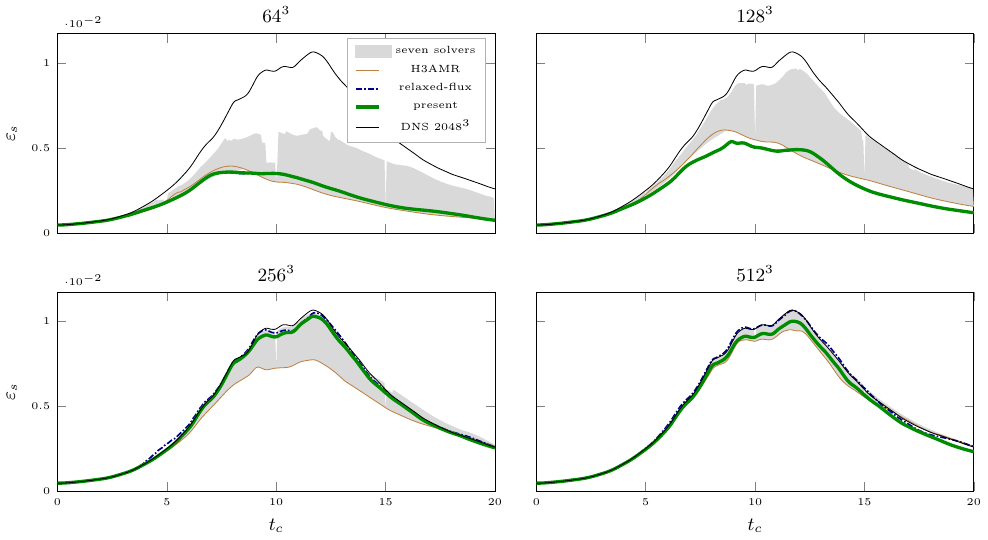}
    \caption{Time evolution of solenoidal dissipation, curves as in Figure~\ref{fig:fneq_ctgv}.}
    \label{fig:fneq_ctgv_epss}
\end{figure*}

\begin{figure*}[tbp]
    \centering
    \begin{subfigure}[b]{0.32\textwidth}\centering
        \includegraphics{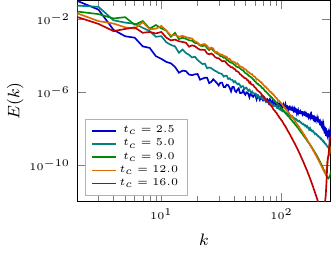}
        \caption{Spectrum through the run}\label{fig:spec_run}
    \end{subfigure}\hfill
    \begin{subfigure}[b]{0.32\textwidth}\centering
        \includegraphics{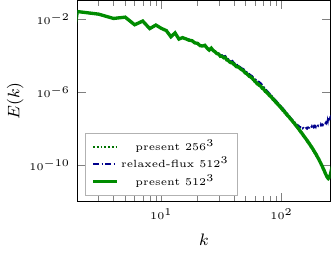}
        \caption{Spectra at $t_c=9$}\label{fig:spec_t9}
    \end{subfigure}\hfill
    \begin{subfigure}[b]{0.32\textwidth}\centering
        \includegraphics{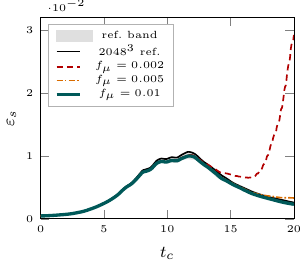}
        \caption{Damping sweep at $512^3$}\label{fig:fmu_sweep}
    \end{subfigure}
    \caption{Small-scale behaviour of the present five-field scheme.
    (a,b) Shell-averaged velocity spectra, normalized so that $\sum_k E(k) = \langle |\mathbf{u}|^2 \rangle / 2$.
    (c) The $512^3$ solenoidal dissipation for three values of the damping $f_\mu$, which separate only in the late tail.}
    \label{fig:fneq_ctgv_spectrum}
\end{figure*}

The five-field scheme lies within the reference band on all three channels at both resolutions (Table~\ref{tab:ctgv_l2}).
The channels are the volume-averaged kinetic energy $E_k$ and the solenoidal and dilatational dissipation rates $\varepsilon_s$ and $\varepsilon_d$, the parts of the viscous dissipation carried by the vorticity and by the dilatation $\nabla\cdot\mathbf{u}$.
At equal resolution the fourteen-field relaxed-flux scheme is more accurate: kinetic energy ties at $256^3$ ($0.0053$), and at $512^3$ the relaxed-flux scheme leads on all three (relaxed-flux/five-field: $E_k$ $0.0028$/$0.0043$, $\varepsilon_s$ $0.013$/$0.053$, $\varepsilon_d$ $0.128$/$0.266$).
Refining from $256^3$ to $512^3$ improves the five-field kinetic energy and dilatational dissipation ($E_k$ from $0.0053$ to $0.0043$, $\varepsilon_d$ from $0.407$ to $0.266$).
The solenoidal dissipation reads a higher error ($\varepsilon_s$ from $0.033$ to $0.053$).
Through the peak of dissipation the five-field curves match the reference to within a few percent at both resolutions (Figures~\ref{fig:fneq_ctgv_epsd} and~\ref{fig:fneq_ctgv_epss}), so the higher $\varepsilon_s$ error is a small uniform under-recovery that stays within the reference band, not a loss of accuracy with refinement.
The $f_\mu$ sweep in Figure~\ref{fig:fneq_ctgv_spectrum}(c) shows the resolved $\varepsilon_s$ is insensitive to $f_\mu$, and the least-diffusive stable setting still leaves the $512^3$ error above the $256^3$ value.
Every other solver in the comparison is high-order, from the fourth-order discontinuous-Galerkin CODA to the sixth-order finite-difference OpenSBLI, and is more accurate per degree of freedom by construction.
H3AMR is the only other second-order scheme and so the like-for-like reference, and against it the five-field scheme is more accurate in every channel at both resolutions.
The reconstruction operator of Section~\ref{sec:reconstruction} departs negligibly from the identity at this working point ($\beta \approx 0.02$).

\begin{table*}[tbp]
    \centering
    \caption{Compressible Taylor--Green vortex, relative $L_2$ error over $t_c\in[0,20]$ against the
    $2048^3$ reference at $256^3$ and $512^3$. The present five-field scheme, its fourteen-field
    baseline and H3AMR are second order, the remaining reference solvers are
    high order.}
    \label{tab:ctgv_l2}
    \begin{tabular}{l ccc ccc}
\toprule
& \multicolumn{3}{c}{$256^3$} & \multicolumn{3}{c}{$512^3$} \\
\cmidrule(lr){2-4}\cmidrule(lr){5-7}
& $E_k$ & $\varepsilon_s$ & $\varepsilon_d$ & $E_k$ & $\varepsilon_s$ & $\varepsilon_d$ \\
\midrule
\multicolumn{7}{@{}l}{\emph{Second order}} \\
    \textbf{Present (5-field)} & 0.0053 & 0.0326 & 0.4068 & 0.0043 & 0.0532 & 0.2660 \\
    Relaxed-flux (14-field)    & 0.0053 & 0.0224 & 0.3370 & 0.0028 & 0.0131 & 0.1278 \\
    H3AMR                      & 0.0119 & 0.2205 & 0.6721 & 0.0059 & 0.0791 & 0.5358 \\
\midrule
\multicolumn{7}{@{}l}{\emph{High order}} \\
    CODA     & 0.0018 & 0.0239 & 0.6009 & 0.0015 & 0.0070 & 0.4289 \\
    OpenSBLI & 0.0035 & 0.0448 & 0.5623 & 0.0006 & 0.0045 & 0.4013 \\
    NS3D     & 0.0029 & 0.0284 & 0.3721 & 0.0009 & 0.0050 & 0.1874 \\
    SD3D     & 0.0136 & 0.0896 & 0.8047 & 0.0041 & 0.0198 & 0.6826 \\
    SPADE    & 0.0076 & 0.0901 & 0.6918 & 0.0040 & 0.0283 & 0.5556 \\
    FLEXI    & 0.0022 & 0.0279 & 0.5095 & 0.0005 & 0.0017 & 0.3282 \\
\bottomrule
\end{tabular}

\end{table*}

\begin{figure*}[tbp]
    \centering
    \resizebox{\textwidth}{!}{\includegraphics{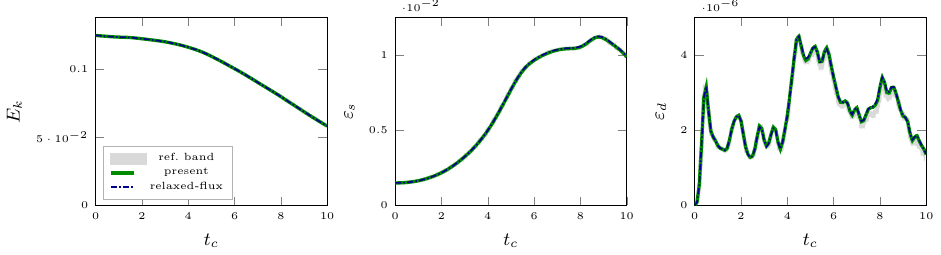}}
    \caption{Subsonic Taylor--Green verification at $\mathrm{Re}=500$, $M_0=0.5$. Green the present
    five-field scheme, blue the fourteen-field relaxed-flux scheme of~\cite{bukreev2026a}, gray the
    envelope of the seven reference solvers~\cite{chapelier2024}.}
    \label{fig:fneq_subsonic}
\end{figure*}

Figure~\ref{fig:fneq_subsonic} repeats the subsonic verification of Chapelier et al.\ at $\mathrm{Re}=500$, $M_0=0.5$, run to $t_c=10$ at $f_\mu=0$ in the shock-free regime.
The present curve tracks the seven-solver band across $E_k$ and the dissipation channels, holding the dilatational peak closer to the band than the relaxed-flux scheme.
The subsonic reference is the envelope digitized from their raster figures, while the supersonic reference of Figure~\ref{fig:fneq_ctgv} is a vector extraction.

\begin{figure*}[tbp]
    \centering
    \includegraphics{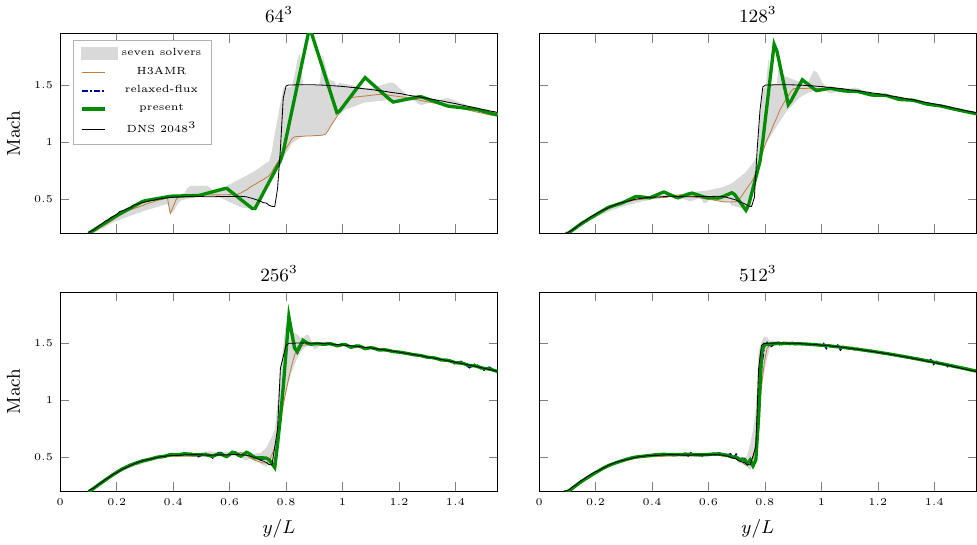}
    \caption{Mach profiles along the $y$ line at $x=z=0$ at $t_c=2.5$, over the first quarter of the
    periodic box. Curves as in Figure~\ref{fig:fneq_ctgv}.}
    \label{fig:fneq_ctgv_mach}
\end{figure*}

The instantaneous fields confirm the integral picture.
Figure~\ref{fig:fneq_ctgv_mach} shows the Mach profile along the $y$ line at $x=z=0$ at $t_c=2.5$, the instant of peak dilatational dissipation.
The five-field profile follows the reference and the fourteen-field relaxed-flux scheme at every resolution and sharpens as the grid refines.
Figure~\ref{fig:fneq_ctgv_spectrum} shows the shell-averaged velocity spectrum.
At equal resolution it matches the fourteen-field relaxed-flux scheme at large scales and rolls off slightly faster at high $k$.
The $512^3$ spectrum broadens through transition and decays after $t_c=9$, and refining from $256^3$ to $512^3$ extends the resolved range without a pile-up at the grid scale, the signature the per-resolution $f_\mu$ is set to avoid.

\section{Performance}\label{sec:performance}
The reconstruction carries five conserved fields instead of the fourteen of the relaxed-flux scheme, forming the velocity and temperature gradients in-register during the collision rather than transporting them as relaxed states.
On memory-bound GPU lattice Boltzmann this is a direct throughput win at accuracy competitive with the relaxed-flux scheme (Section~\ref{sec:ctgv}).

\paragraph{Memory traffic}
Both schemes are D3Q7, so each field streams seven populations per cell.
The relaxed-flux scheme carries fourteen fields, $14\times7=98$ populations and \qty{784}{\byte} per cell in single precision counting the read and the write.
The five-field scheme carries $5\times7=35$ populations at \qty{280}{\byte} and reads the per-cell shock-sensor field, \qty{284}{\byte} per cell in total, a $2.8\times$ reduction.
The reconstructed gradients are formed in registers and never stored, so they add no traffic.
Since GPU lattice Boltzmann is memory-bound, this ratio sets the throughput ceiling.

\begin{table*}[tbp]
\centering
\caption{Performance and per-kernel resource characteristics on an NVIDIA RTX A5000
(\qty{680}{\giga\byte\per\second} peak bandwidth measured with BabelStream Triad~\cite{deakin2018}, \qty{27.8}{\tera\flop\per\second} FP32), single
precision, D3Q7, at $256^3$, timed on the solver alone.
\emph{AI} is arithmetic intensity in FLOP/byte, \emph{BW} the achieved bandwidth,
\emph{BW sat.}\ its fraction of peak, and \emph{Occ.}\ the register-limited occupancy of the collide
kernel. \textsuperscript{c}\,operation count of the executed collide after
\emph{Common-Subexpression Elimination} (CSE), as in~\cite{kummerlaender2026d}.}
\label{tab:fneq_roofline}
\resizebox{\textwidth}{!}{%
\begin{tabular}{l c c c c c c c c c c c}
\toprule
\textbf{Scheme} & \textbf{Fields} & \textbf{FLOP/cell} & \textbf{Byte/cell} & \textbf{AI} & \textbf{MLUP/s} & \textbf{BW} & \textbf{BW sat.} & \textbf{GFLOP/s} & \textbf{Reg.} & \textbf{Spill} & \textbf{Occ.} \\
 & & & & & & [\unit{\giga\byte\per\second}] & & & & [\unit{\byte}] & \\
\midrule
Reconstruction & 5  & 1080\textsuperscript{c} & 284 & 3.80 & 2028 & 576 & 85\% & 2190 & 128 & 0 & 33 \\
Relaxed-flux    & 14 & 1864\textsuperscript{c} & 784 & 2.38 & 413  & 323 & 48\% & 769  & 255 & 0 & 17 \\
\bottomrule
\end{tabular}}
\end{table*}

\begin{figure}[tbp]
    \centering
    \includegraphics[width=\columnwidth]{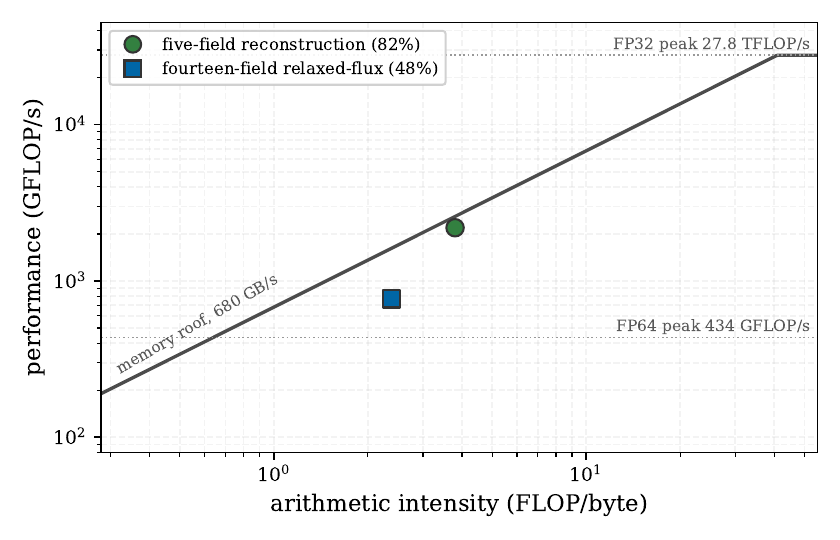}
    \caption{Roofline on the RTX A5000 (Table~\ref{tab:fneq_roofline}). Both kernels lie in the
    memory-bound region, left of the ridge at $I=40.9$. Reducing fourteen fields to five moves the
    kernel toward the roof.}
    \label{fig:fneq_roofline_plot}
\end{figure}

\paragraph{Throughput and the roofline}
Table~\ref{tab:fneq_roofline} collects the measured throughput in \emph{Million Lattice Updates per Second} (MLUP/s) and Figure~\ref{fig:fneq_roofline_plot} places the kernels on the roofline.
At $256^3$, single precision, the five-field scheme sustains \qty{2028}{\mega\lups} against the relaxed-flux scheme's \qty{413}{\mega\lups} on the same GPU, a $4.9\times$ speed-up, above the $2.8\times$ traffic ratio because the relaxed-flux scheme does not saturate memory.
Each kernel is built at its throughput optimum and neither spills: the five-field collide packs into 128 registers for two blocks per streaming multiprocessor, while the fourteen-field collide needs all 255 and runs one.
The five-field kernel reaches $85\%$ of peak bandwidth, the relaxed-flux scheme only $48\%$.
The relaxed-flux scheme moves more than twice the bytes per cell yet converts fewer of them to throughput.
Forming and relaxing fourteen moments per cell, against the reconstruction's five, leaves its collide unable to saturate the memory system.

\paragraph{Time to solution and problem size}
Efficiency is ultimately time to a target accuracy and problem size, not raw throughput.
At $256^3$ the two schemes are of comparable accuracy (Table~\ref{tab:ctgv_l2}), so the wall-clock ratio is the $4.9\times$ throughput ratio on the same hardware.
The $2.8\times$ smaller footprint is equally consequential for the finest grid that fits in a given memory budget.
At $512^3$ the five-field state is \qty{19}{\giga\byte} and runs on the two-GPU node used here, whereas the fourteen-field state is \qty{53}{\giga\byte} and does not fit, so the five-field scheme reaches grid resolutions on a given machine that the relaxed-flux scheme cannot.
The relaxed-flux scheme is more accurate per grid point in the dissipation channels where memory and time do not bind.
The reconstruction is the efficient choice where they do, trading a modest dissipation-accuracy margin for $2.8\times$ less memory and several times the throughput.

\section{Conclusion}
We presented a lattice Boltzmann method for the compressible Navier--Stokes--Fourier equations that transports only the conserved state.
The velocity and temperature gradients the Newtonian and Fourier closures need are recovered inside the collision from the conserved fields' non-equilibrium first moments, by the reconstruction operator of~\cite{kummerlaender2026e}, so the viscous stress and heat flux are no longer transported as auxiliary state.
The scheme carries five fields on the D3Q7 lattice where the relaxed-flux form of~\cite{bukreev2026a} carries fourteen, and reaches supersonic shocklet-containing turbulence on a compact local lattice with exact streaming, entirely in single precision.

On the exact Sod and Becker solutions the five-field scheme matches the fourteen-field relaxed-flux baseline.
On the compressible Taylor--Green vortex at $\mathrm{Re}=1600$, $M_0=1.25$ it stays within the seven-solver reference band of~\cite{chapelier2024} on the kinetic energy and both dissipation rates at $256^3$ and $512^3$, and against H3AMR, the only other second-order scheme, it is more accurate in every channel at both resolutions.

Carrying five fields rather than fourteen moves $284$ bytes per cell against $784$, a $2.8\times$ reduction, and on an RTX A5000 the collide sustains $2028$ against $413$ million lattice updates per second in single precision, a $4.9\times$ speed-up at $85\%$ of peak bandwidth.
The smaller footprint is as consequential as the speed, as the $512^3$ five-field state is \qty{19}{\giga\byte} and runs on the two-GPU node used here where the fourteen-field state at \qty{53}{\giga\byte} does not fit.
The relaxed-flux scheme remains more accurate per grid point in the dissipation channels, and the reconstruction is the efficient choice wherever memory and time bind.

The CNSF system thus joins those solved by reconstruction from a single declaration~\cite{kummerlaender2026d, kummerlaender2026e}, rather than by transporting auxiliary state.

\section*{Code Availability}

The five-field scheme is fully specified in this paper: the declaration of Listing~\ref{lst:declaration}, the reconstruction operator of Section~\ref{sec:cnsf_operator}, the shock-capturing collision of Listing~\ref{lst:shocksensor}, and the benchmark configurations of Section~\ref{sec:validation}.
Its generated collide-and-stream kernels run in the open-source framework OpenLB~\cite{krause2021}, release 1.9~\cite{kummerlaender2025f}.
The \emph{PDE2LBM} compiler and the simulation cases are available upon reasonable request.

\section*{Acknowledgments}

Google Gemini and Anthropic Claude assisted in drafting and revising the manuscript.
All content was reviewed, verified, and approved by the authors, who take full responsibility for it.

\bibliographystyle{elsarticle-num}
\bibliography{literature}

\end{document}